\documentclass[12pt]{article}
\usepackage{pic03}
\usepackage{hyperref}
\usepackage{url}
\usepackage{graphicx}

\begin{document}

\newcommand{\lesssim}{\stackrel{\textstyle <}{\raisebox{-.6ex}{$\sim$}}}
\newcommand{\gtrsim}{\stackrel{\textstyle >}{\raisebox{-.6ex}{$\sim$}}}

\title{\bf NEUTRINO OSCILLATIONS: A GLOBAL ANALYSIS}
\author{
G.L. Fogli, E. Lisi, A. Marrone, D. Montanino$^*$,  A. Palazzo, A.M. Rotunno      \\
{\em Dipartimento di Fisica \& Sezione INFN, Bari}\\
{\em $^*$Dipartimento di Fisica \& Sezione INFN, Lecce}}
\maketitle

%
%
%
%
%
%
\vspace{4.0cm}
%

\baselineskip=14.5pt
\begin{abstract}
We review the status of the neutrino oscillation physics (as of
June 2003), with a particular emphasis on the present knowledge of
the neutrino mass-mixing parameters in a three generation
approach. We consider first the $\nu_\mu\to\nu_\tau$ flavor
transitions of atmospheric neutrinos. It is found that standard
oscillations provide the best description of the SK+K2K data, and
that the associated mass-mixing parameters are determined at
$\pm1\sigma$ (and $N_\mathrm{DF}=1$)  as: $\Delta m^2=(2.6\pm
0.4)\times 10^{-3}$ eV$^2$ and $\sin^2
2\theta=1.00^{+0.00}_{-0.05}$. Such indications, presently
dominated by SK, could be strengthened by further K2K data. Then
we analyze the energy spectrum of reactor $\nu$ events recently
observed at KamLAND and combine them with solar and  terrestrial
$\nu$ data. We find that the solution to the solar $\nu$ problem
at large mixing angle (LMA) is basically split into two
sub-regions, that we denote as LMA-I and LMA-II. The LMA-I
solution, characterized by lower values of the squared neutrino
mass gap, is favored by the global data fit. Finally, we briefly
illustrate how prospective data from the SNO and KamLAND can
increase our confidence in the occurrence of standard matter
effects in the Sun, which are starting to emerge from current
data.\end{abstract}
\newpage

\baselineskip=17pt

\section{Introduction}
In its first phase of operation (years 1996--2001), the
Super-Kamiokande (SK) experiment has provided, among other
important results, compelling evidence for atmospheric $\nu_\mu$
disappearance \cite{Evid,KaTo}. This evidence, now firmly based on
a high-statistics 92 kton-year exposure \cite{Sh02}, has not only
been corroborated by consistent indications in the
MACRO \cite{MACR} and Soudan~2 \cite{Soud} atmospheric neutrino
experiments, but has also been independently checked by the first
long-baseline KEK-to-Kamioka (K2K) accelerator experiment
\cite{K2K1,K2K2}, using SK as a target for $\nu_\mu$'s produced
250 km away with $\langle E_\nu\rangle \sim 1.3$~GeV.
Neutrino flavor oscillations, interpreted in terms of nonzero mass-mixing
parameters ($\Delta m^2,\sin^2 2\theta$) in the
$\nu_\mu\to\nu_\tau$ channel, provide by far the best and most
natural explanation for the observed
$\nu_\mu$ disappearance \cite{Evid,KaTo}.

In Section 2 we review the phenomenological status of the
standard oscillations in the
$\nu_\mu\to\nu_\tau$ channel, in the light of the latest SK
atmospheric zenith distributions \cite{Sh02}
and of the first spectral results from the K2K experiment
\cite{K2K2}.

On the solar neutrino front,
the Chlorine \cite{Cl98}, Gallium \cite{Ab02,Ha99,Ki02},
Super-Kamiokande (SK) \cite{Fu01,Fu02} and Sudbury Neutrino
Observatory (SNO) \cite{AhCC,AhNC,AhDN} solar neutrino experiments
have convincingly established that the deficit of the observed
solar $\nu_e$ flux with respect to expectations \cite{BP00}
implies new neutrino physics. In particular, the charged and
neutral current (CC and NC) data from SNO have proven the
occurrence of $\nu_e$ transitions into a different active state
$\nu_a$ with a statistical significance greater than $5\sigma$
\cite{AhNC}.

Barring sterile neutrinos and nonstandard $\nu$ interactions, such
transitions can be naturally explained by the hypothesis of flavor
oscillations \cite{Pont} in the $\nu_e\to\nu_a$ channel ($\nu_a$
being a linear combination of $\nu_\mu$ and $\nu_\tau$) driven by
nonzero $\nu$ squared mass difference and mixing angle parameters
($\delta m^2, \theta_{12}$) \cite{Maki}. The $(\nu_\mu,\nu_\tau)$
combination orthogonal to $\nu_a$ is probed by atmospheric $\nu$
oscillations \cite{KaTo}, with different parameters $(\Delta
m^2,\theta_{23})$ \cite{Revi}. The third mixing angle
$\theta_{13}$, needed to complete the $3\times 3$ mixing matrix,
is constrained to be small by additional reactor results
\cite{CHOO,Palo}, and can be set to zero to a good
approximation for our purposes.

The recent results from the Kamioka Liquid scintillator
AntiNeutrino Detector (KamLAND) \cite{KamL} have provided a
beautiful and crucial confirmation of the solar $\nu_e$
oscillation picture through a search for long-baseline
oscillations of reactor $\overline\nu_e$'s. The observed of
$\overline\nu_e$ disappearance in KamLAND has confirmed the
previously favored solution in the $(\delta m^2,\theta_{12})$
parameter space, often referred to as the large mixing
angle (LMA) region \cite{AhDN} in the literature (see, e.g.,
\cite{Last} and references therein). Moreover, the KamLAND data
have basically split this region into two allowed subregions,
which we will refer to as LMA-I and LMA-II, following
Ref.~\cite{KLou}.

In Sections 3 we analyze the first KamLAND spectral data
\cite{KamL} and combine them with current solar neutrino data
\cite{Cl98,Ab02,Ha99,Ki02,Fu02,AhCC,AhNC,AhDN}, assuming two-flavor
oscillations of
active neutrinos \cite{Last}, in order to determine the surviving
sub-regions of the
LMA solution. In the analysis we include the CHOOZ reactor data
\cite{CHOO}.

Finally, in Section 4 we  briefly illustrate how
emerging indications of solar matter effects can be corroborated
in the LMA parameter region. In particular, we show that the
amplitude of matter effects (introduced as a free parameter
$a_\mathrm{MSW}$) can be significantly constrained by
using prospective data from SNO and KamLAND.
%
%
%
\section {``Atmospheric" neutrinos}
%

A careful analysis of the SK and K2K data sets
used in the following can be found in \cite{atm0}.
Concerning SK atmospheric neutrino data (92 kton-year
\cite{Sh02}), we use the usual zenith angle
$(\theta_z)$ distributions of leptons: sub-GeV $e$-like and
$\mu$-like events, divided in 10+10 bins;
multi-GeV $e$-like and $\mu$-like events,
divided in 10+10 bins; upward stopping and through-going $\mu$
events, divided in 5+10 bins. The
calculation of the theoretical events rates $R_n^\mathrm{theo}$ in
each of the 55 bins is done as in \cite{Fo98,Fo01,Ma01}. The SK
statistical analysis is considerably improved with respect to
\cite{Fo98,Ma01}. Now the set of
systematic errors has been enlarged to 11 entries, leading to a
more complex structure of correlated errors affecting the
$R_n^\mathrm{theo}$'s, as emphasized in
\cite{GetM}.

Concerning the K2K data, we use the absolute spectrum of muon
events in terms of the reconstructed neutrino energy $E$
\cite{K2K2}, which provides a total of 29 events (here
divided in 6 bins). In this sample, the parent neutrino
interactions are dominantly quasi-elastic (QE), and the
reconstructed energy $E$ is thus closely correlated with the true
neutrino energy $E_\nu$.

Let us now discuss the updated bounds on the parameters $(\Delta
m^2,\,\sin^2 2\theta)$, governing the scenario of standard
oscillations (here $\theta=\theta_{23}$).

%
\begin{figure}[t]
\includegraphics[scale=0.850]{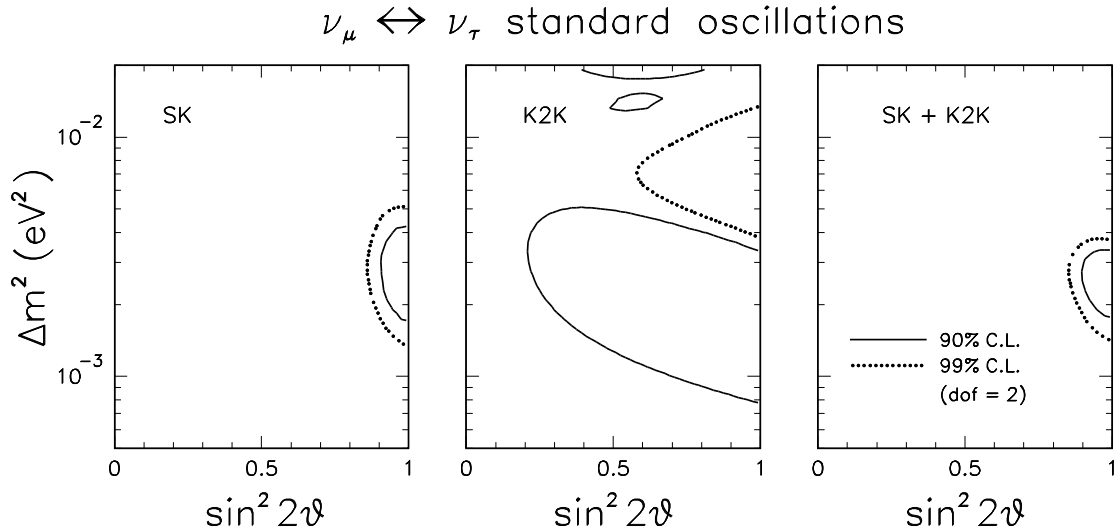}
\caption{\it Standard oscillations in the $\nu_\mu\to\nu_\tau$
channel: bounds on the parameters $(\Delta m^2,\,\sin^2 2\theta)$
from SK atmospheric data (left panel), K2K spectral data (middle
panel), and their combination (right panel).}
\end{figure}



Fig. 1 shows the joint bounds on the $(\Delta
m^2,\,\sin^2 2\theta)$ parameters from our analysis of SK, K2K,
and SK+K2K data.
The bounds in the left panel are very close
to the official SK ones, as presented in \cite{Sh02}.
The bounds in the middle panel are instead slightly weaker than
the official K2K ones \cite{K2K2}, especially in terms of $\sin^2
2\theta$. In particular, we do not find a lower bound on
$\sin^22\theta$ at 99\% C.L.\ (for $N_\mathrm{DF}=2$). The reason
is that we cannot use the additional (dominantly) non-QE event
sample of K2K (27 events), which would help to constrain the
overall rate normalization and thus $\sin^2 2\theta$. This fact
might also explain why we find the K2K best  fit at $\sin^2 2\theta=0.82$
rather than at 1.00 as in \cite{K2K2}.
By comparing the left and right panels of Fig.~1, the
main effect of K2K appears to be the strengthening of the upper
bound on $\Delta m^2$, consistently with the trend of the first
K2K data (rate only \cite{K2K1}, no spectrum) \cite{Ma01}. The
main reason is that, for $\Delta m^2\sim (4$--$6)\times 10^{-3}$
eV$^2$, the first oscillation minimum would be located at---or
just above---the K2K energy spectrum peak, implying a strong local
and overall suppression of the expected events.

%
%
\begin{figure}[htbp]
  \centerline{\hbox{ \hspace{0.1cm}
    \includegraphics[width=6.5cm]{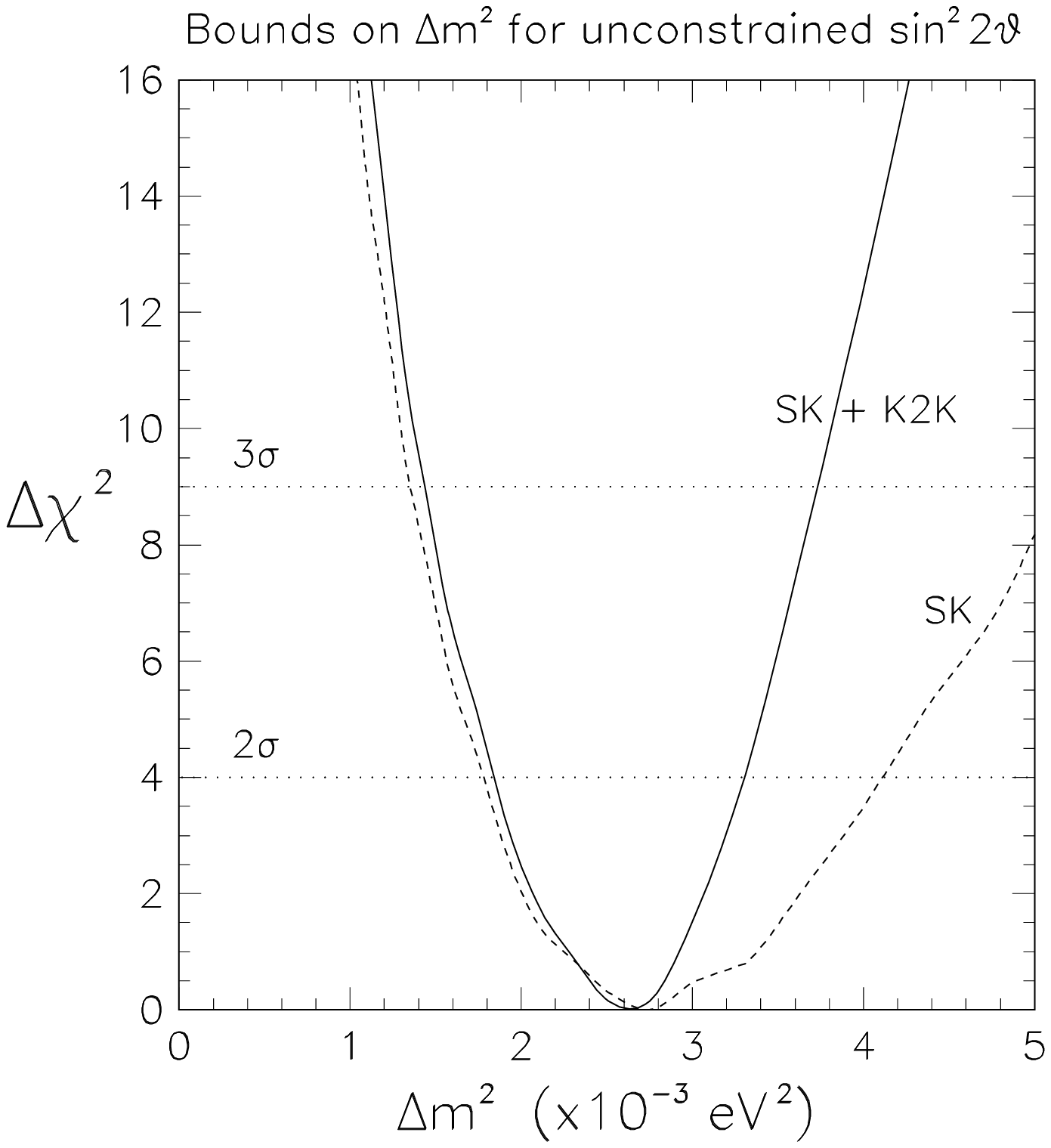}
    \hspace{0.3cm}
    \includegraphics[width=6.5cm]{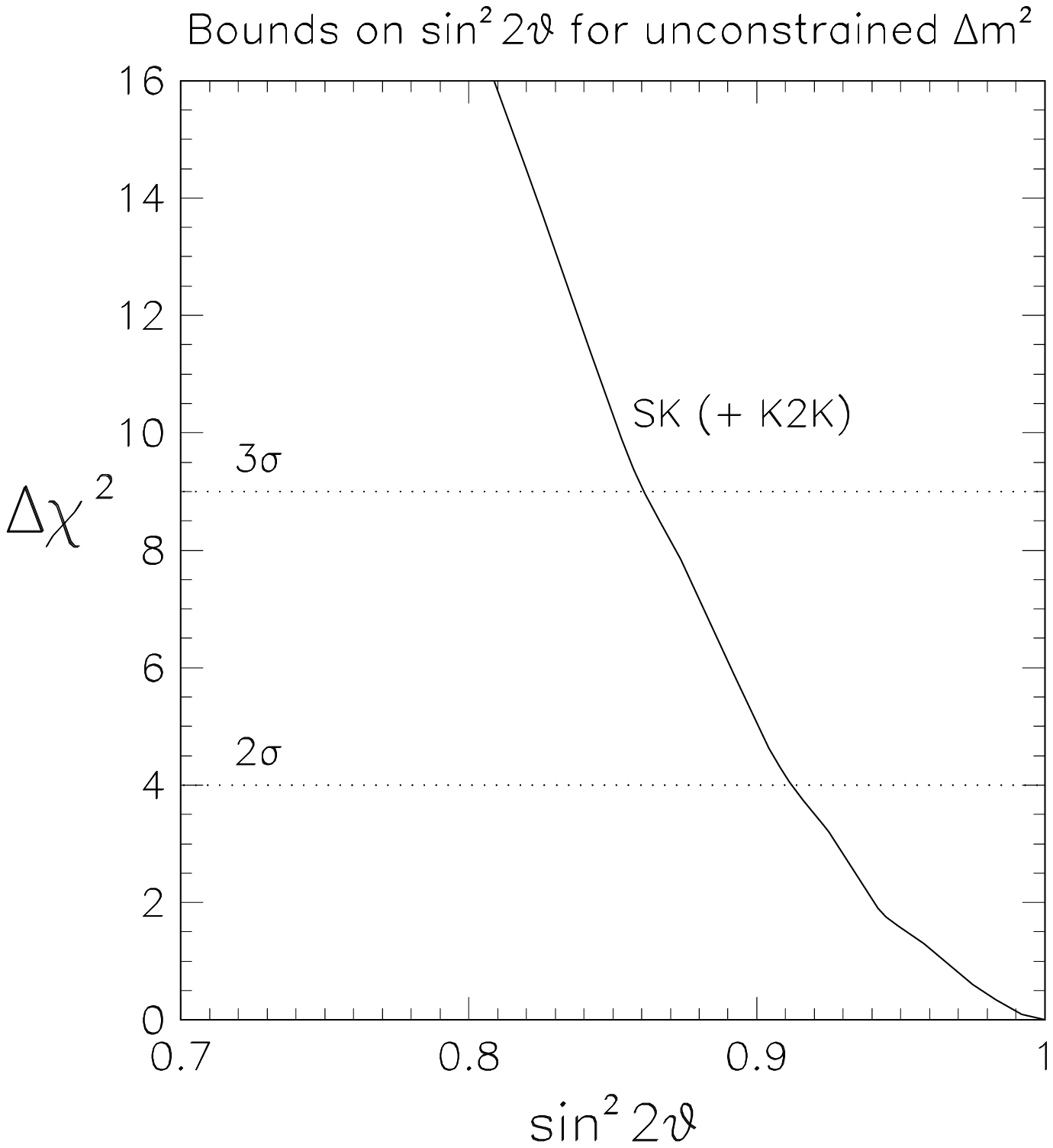}
    }
  }
 \caption{\it
      Standard oscillations in the
$\nu_\mu\to\nu_\tau$ channel. On the left: bounds on $\Delta m^2$ for
unconstrained $\sin^22\theta$ from SK (dashed curve) and SK+K2K
(solid curve). On the right: bounds on $\sin^2 2\theta$ for
unconstrained $\Delta m^2$ from SK data. The inclusion of K2K data
induces here negligible changes (not shown).
    \label{twofig} }
\end{figure}

Fig.~2 shows on the left the SK and SK+K2K bounds on $\Delta m^2$,
when the $\sin^2 2\theta$ parameter is projected (minimized) away.
The linear scale in $\Delta m^2$ makes the K2K impact on
the upper limit more evident. Notice that, up to $\sim\!3\sigma$, the global
(SK+K2K)  $\chi^2$ function
is approximately parabolic in the {\em
linear\/} variable $\Delta m^2$, so that one can define a one-
standard-deviation error for this parameter. This
feature was already argued on the basis of a
graphical reduction of the official SK and K2K likelihood
functions \cite{Last}, and is here confirmed through a full
analysis. By keeping only the first significant figure in the
error estimate, a parabolic fit provides the $\pm 1\sigma$ range
\begin{equation}
\label{Dm2range}
 \Delta m^2 = (2.6 \pm 0.4)\times 10^{-3}\mathrm{\ eV}^2\ .
\end{equation}

The bounds on $\sin^2 2\theta$ are instead entirely dominated by
SK. This is shown on the right of Fig.~3, where the  $\Delta\chi^2$ function in terms
of $\sin^2 2\theta$ is reported, for $\Delta m^2$ projected (minimized) away
in the SK fit. Here the addition of K2K data would
insignificantly change the bounds (not
shown), which thus hold for
both the SK and the SK+K2K fit. Also in this case, the nearly
parabolic behavior of $\Delta \chi^2$ allows to properly define a
$1\sigma$ range,

\begin{equation}
\label{sinrange} \sin^2 2\theta = 1.00^{+0.00}_{-0.05} \ ,
\end{equation}
with the lower $N\sigma$ error scaling linearly with $N$ (up to
$N\simeq 3$). Equations~(\ref{Dm2range}) and (\ref{sinrange})
concisely review the current fit to the standard oscillation
parameters, as anticipated in the Introduction.

%
\begin{figure}[t]
\vspace{0pt}
\hspace{20pt}
\includegraphics[scale=0.8]{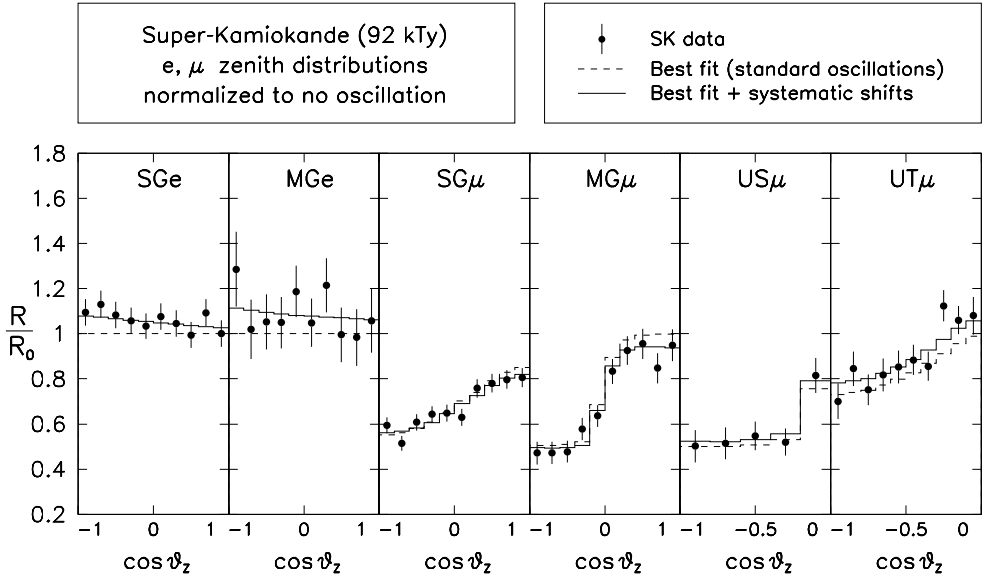}
\vspace{-0pt}
\caption{\it
SK experimental zenith distributions
($\pm 1\sigma_\mathrm{stat}$), compared with the corresponding
theoretical ones at the global (SK+K2K) best-fit point.
All distributions are normalized to the unoscillated predictions
in each bin. For the
theoretical event rates, we show both the central values $R_n^\mathrm{theo}$
(dashed histograms) and the ``shifted'' values $\overline
R_n^\mathrm{theo}$ (solid histograms), which embed the effect of
systematic pulls. The difference between $\overline
R_n^\mathrm{theo}$ and $R_n^\mathrm{theo}$ shows how much (and in
which direction) the correlated systematic errors tend to stretch
the predictions in order to match the data.}
\end{figure}


Fig.~3 shows the comparison between observations and
best-fit predictions for the
SK zenith distributions.
In particular, the comparison between solid and dashed
histograms shows that the systematic shifts are often comparable
in size to the statistical errors, implying that just increasing
the SK atmospheric $\nu$ statistics will hardly bring decisive new
information on the standard oscillation scenario.
In the SG and MG samples, the fit clearly exploits the
systematic uncertainties to increase the $e$-like event
normalization, especially in the upward direction, so as to
reduce the ``electron excess'' possibly indicated by SK data.

Concerning $\mu$-like events in the
SG and MG samples, the fit shows an opposite tendency to slightly
decrease the normalization of (especially down-going) events. The
tendency appears to be reversed in the high-energy UT sample.
Taken together, these opposite shifts of $e$-like and $\mu$-like
expectations in the SG and MG samples seem to suggest some
systematic deviation from the predicted $\mu/e$ flavor ratio
which, although not statistically alarming, should be kept in mind:
deviations
of similar size might have their origin in neutrino physics beyond
$2\nu$ oscillations. Unfortunately, since such effects are
typically not larger than the systematic shifts in
Fig.~3, they are likely to remain hidden in
higher-statistics SK data.

\section{Impact of KamLAND on solar neutrinos (a $2\nu$ analysis)}
%

%
\begin{figure}[t]
\vspace{0pt}
\hspace{70pt}
\includegraphics[scale=0.5]{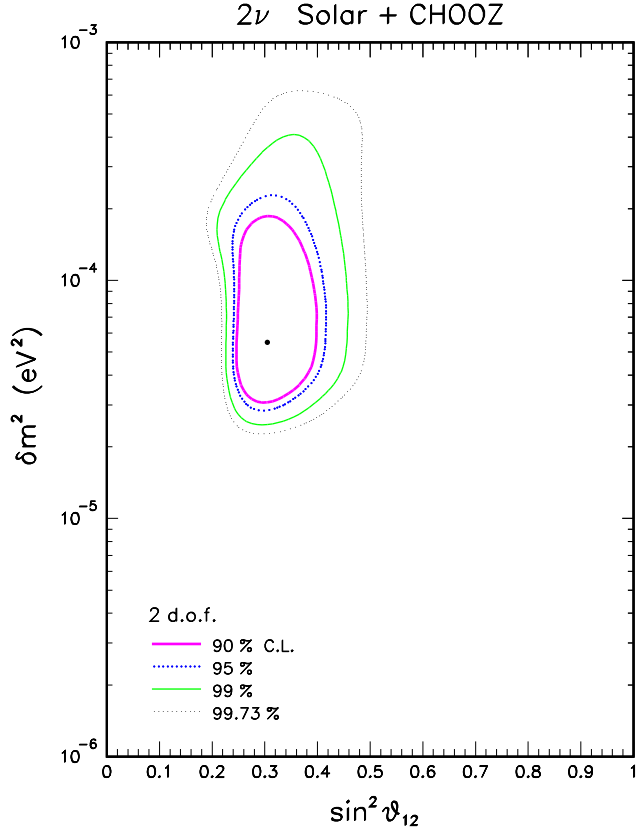}
\vspace{-5pt}
\caption{\it
Two-flavor active neutrino
oscillations: Global analysis of solar and CHOOZ neutrino data in
the $(\delta m^2,\sin^2\theta_{12})$ parameter space, restricted
to the LMA region. The best fit is indicated by a black dot.}
\end{figure}


\noindent
The KamLAND recent observation of $\overline\nu_e$ disappearance
\cite{KamL} confirms the current interpretation of solar neutrino
data \cite{Fu02,AhDN,Last,GetM} in terms of
$\nu_e\to\nu_{\mu,\tau}$ oscillations induced by neutrino mass and
mixing \cite{Pont,Maki}, and restricts the corresponding parameter
space $(\delta m^2,\theta_{12})$ within the so-called large mixing
angle (LMA) region. In this region, globally favored by solar
neutrino data, matter effects \cite{Matt,Barg} in
adiabatic regime \cite{Adia,Matt} are expected to dominate the
dynamics of flavor transitions in the Sun (see, e.g.,
\cite{LasA}). The KamLAND spectral data appear to exclude some
significant portions of the LMA solution \cite{KamL}, where the
predicted spectrum distortions
\cite{Last} would be
in conflict with observations \cite{KamL}.

In the $2\nu$ case, we find that the inclusion of the KamLAND
spectrum basically splits the LMA solution into two sub-regions at
``small'' and ``large'' $\delta m^2$, which we call LMA-I and
LMA-II, respectively (the LMA-I solution being preferred by the
data) \cite{sol0}. Such regions are only slightly modified in the
presence of $3\nu$ mixing, namely, for nonzero values of the
mixing angle $\theta_{13}$. We also present updated bounds (as of
June 2003) in the $3\nu$ parameter space $(\delta
m^2,\theta_{12},\theta_{13})$.

In our KamLAND data analysis \cite{sol0}, we use the absolute spectrum of
events reported in \cite{KamL}, taken above a background-safe
analysis threshold of 2.6 MeV in visible energy $E$.
The events below such threshold might contain a significant
component of geological $\overline\nu_e$'s \cite{GeoN}, whose
large normalization uncertainties are poorly constrained at
present by the KamLAND data themselves \cite{KamL}. Above 2.6 MeV,
a total of 54 events is found, against 86.8 events expected from reactors
\cite{KamL}. Finally, the observed energy spectrum of events is analyzed as in
\cite{Last}, with the improvements reported in \cite{sol0}.

An updated $2\nu$ analysis of current solar+CHOOZ neutrino data,
\cite{Last}, is presented here. The fit includes 81 solar neutrino observables
\cite{GetM,Last} and 14 CHOOZ spectrum bins \cite{CHOO,Last}, for
a total of 95 data points. The $\Delta\chi^2$ expansion around
the minimum, relevant
for the estimation of the oscillation parameters $(\delta
m^2,\sin^2\theta_{12})$, is shown in Fig.~4, where we have
restricted the $\delta m^2$ range to the only three decades
relevant for the LMA solution and for the following KamLAND
analysis.

%
\begin{figure}[t]
\hspace{75pt}
\includegraphics[scale=0.5]{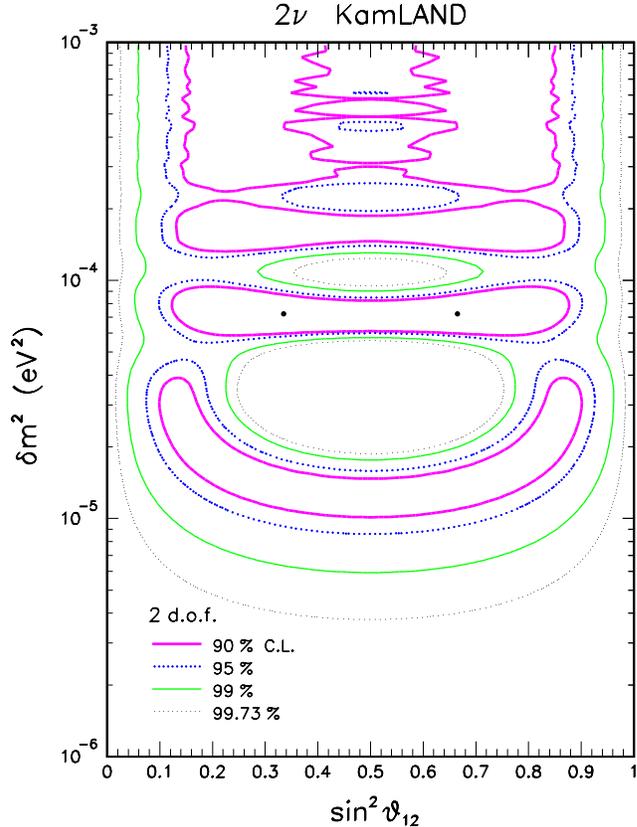}
\caption{\it Two-flavor active neutrino oscillations: Analysis of
the KamLAND energy spectrum data above 2.6 MeV in the $(\delta
m^2,\sin^2\theta_{12})$ parameter space. A ``tower'' of
octant-symmetric regions is allowed at different values of $\delta
m^2$. The symmetric best fits are indicated by black dots. The
left dot is remarkably close to the solar best fit in Fig.~4.}
\end{figure}



In Fig.~5 we report the $2\nu$ analysis of KamLAND \cite{sol0}:
there appears to be a ``tower'' of solutions which tend to merge
and become indistinguishable for increasing $\delta m^2$; the
lower three ones are, however, rather well separated at 90\% C.L.
Notice that our allowed regions are slightly larger (i.e., less
constraining) than those in the rate+shape analysis of
\cite{KamL}.  One of the two
octant-symmetric best fits points in Fig.~5 (black dots) is
remarkably close to the best fit in Fig.~4. The
difference in location with respect to the KamLAND official
best-fit point at $(\delta
m^2/\mathrm{eV}^2,\sin^2\theta_{12})=(6.9\times 10^{-5},\,0.5)$
\cite{KamL} is not statistically significant, amounting to a
variation $\Delta\chi^2\ll 1$.

%
\begin{figure}[t]
\vspace{-10pt}
\hspace{70pt}
\includegraphics[scale=0.5]{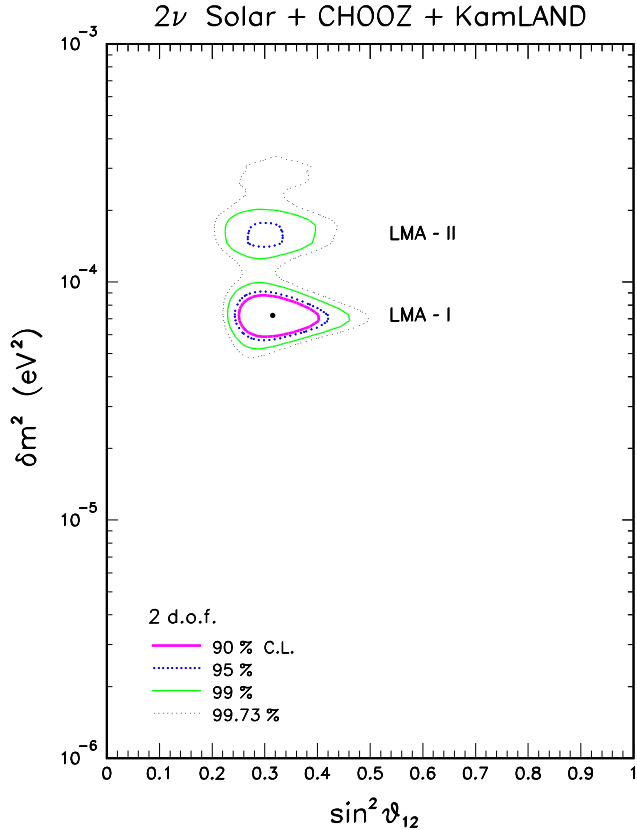}
\vspace{-5pt} \caption{\it Two-flavor active neutrino
oscillations: Global analysis of solar, CHOOZ, and KamLAND
neutrino data in the $(\delta m^2,\sin^2\theta_{12})$ parameter
space. With respect to Fig.~4, the LMA region is significantly
restricted, and appears to be split into two sub-regions (LMA-I
and LMA-II), well-separated at 99\% C.L.}
\end{figure}


The combination of the solar+CHOOZ results in Fig.~4 with the
KamLAND results in Fig.~5 gives the global $2\nu$ results shown in
Fig.~6. Two rather
distinct solutions, that we label LMA-I and LMA-II, are seen to
emerge. The LMA-I solution is clearly preferred by the data,
being close to the best fit points of both solar+CHOOZ and KamLAND
data. The LMA-II solution is located at a $\delta m^2$ value about
twice as large as for the LMA-I, but is separated from the latter
by a modest $\Delta\chi^2=5.4$ difference (dominated by solar
neutrino data). Indeed, if we conservatively demand a 99.73\%
C.L.\ for the allowed regions, the LMA-I and LMA-II solutions
appear to merge (and extend towards $\delta m^2\sim 3\times
10^{-4}$ eV$^2$) in a single broad solution. In any case, at any
chosen C.L., the allowed regions of Fig.~6 are significantly
smaller than those in Fig.~4. Therefore, with just 54 initial
events, the KamLAND experiment is not only able to select the LMA
region as {\em the\/} solution to the solar neutrino problem, but
can also significantly restrict the corresponding oscillation
parameter space. With several hundred events expected in the
forthcoming years, there are thus very good prospects to refine
the parameter estimate \cite{Last}.

\section{Indications of matter effects in the Sun}

Within the LMA region, solar neutrino oscillations are governed
not only by the kinematical mass-mixing parameters $(\delta m^2,
\theta_{12}$), but should also be significantly affected by the
interaction energy difference ($V=V_e-V_a$) between $\nu_e$'s and
$\nu_a$'s propagating in the solar (and possibly Earth) background
matter \cite{Matt,Barg}, through the so-called
Mikheev-Smirnov-Wolfenstein (MSW) mechanism \cite{Matt} in
adiabatic regime \cite{Adia}. Although Earth matter effects (i.e.,
day-night variations of solar event rates) remain elusive, solar
matter effects seem to emerge, at least indirectly, from the
combination of the available data (and especially from SNO),
through a preference for an average oscillation probability
smaller than $1/2$ at energies of a few MeV.
A phenomenological approach to the problem has been recently presented
in \cite{LasA}, where a
free parameter $a_\mathrm{MSW}$ is introduced, called to modulate the overall
amplitude of the interaction energy difference $V$ in the dynamical term
${\cal H}_\mathrm{dyn}$ of the Hamiltonian,
\begin{equation}
\label{Ha} V \to a_\mathrm{MSW}\cdot V\ .
\end{equation}
By treating $a_\mathrm{MSW}$ as a continuous parameter, one can
try to constrain its allowed range through global data analyses: A
preference for $a_\mathrm{MSW}\sim O(1)$ would then provide an
indirect indication for the occurrence of matter effects with
standard size, as opposed to the case of pure ``vacuum''
oscillations (${\cal H}_\mathrm{dyn}\simeq 0$).%

We have verified that the current solar neutrino data, by
themselves, place only very loose and uninteresting limits on
$a_\mathrm{MSW}$, as far as the mass-mixing oscillation parameters
are left unconstrained. In fact, since the oscillation physics
depends mostly on the ratio $V/k$, where $k=\delta
m^2/2E$ is the neutrino oscillation wavenumber,
a variation of the kind $V\to
a_\mathrm{MSW}V$ is largely absorbed by a similar rescaling of $k$
(i.e., of $\delta m^2$). In order to break this degeneracy, we
need to include explicitly an experiment which is highly sensitive
to $\delta m^2$ and basically
insensitive to matter effects, such as KamLAND.

A SSM-independent
preference for $P_{ee}<1/2$ has been provided first by the
combination of SNO CC and SK data \cite{AhCC} and then by SNO data
alone through the CC/NC double ratio \cite{AhNC}, but not yet with
a significance high enough to rule out $P_{ee}=1/2$ \cite{GetM}.
Let us consider, in particular, the latest SNO constraints in the
plane $(\Phi_e,\Phi_{\mu\tau})$ charted by the solar $\nu_e$ and
$\nu_{\mu,\tau}$ fluxes, as shown in Fig.~3 of the original SNO
paper \cite{AhNC}. In such a figure, although the SNO best-fit
point clearly prefers $P_{ee}\sim 1/3$ (corresponding to
 $\Phi_{\mu\tau}\simeq 2\Phi_e$), the 95\% C.L.\
ellipse is still compatible with $P_{ee}\sim 1/2 $ (namely,
$\Phi_{\mu\tau}\simeq \Phi_e$). However, future SNO NC and CC data
 can considerably improve the constraints on $P_{ee}$, by
reducing both the statistical and the systematic error on the
CC/NC ratio \cite{Hall}.

In conclusion, although the combination of all current solar
neutrino data suggests a pattern of $P_{ee}$ compatible with the
LMA energy profile and indicates an overall
preference for the first octant of $\theta_{12}$ \cite{AhDN}, the
emerging indications in favor of solar matter effects from this
data set are not strongly compelling yet.

Until now we have illustrated how a single
datum (the SNO CC/NC double ratio) can discriminate the case of
standard matter effects $(a_\mathrm{MSW}=1)$ from the case of
zeroed matter effects $(a_\mathrm{MSW}=0)$ in the LMA parameter
region. By using further experimental information from KamLAND,
one could try to test \cite{LasA} whether the ``solar~+~KamLAND'' combination
of data can constrain matter effects in the Sun to have the right
size [$a_\mathrm{MSW}\sim O(1)$]. In this kind of analyses,
KamLAND basically fixes the oscillation parameters $(\delta
m^2,\sin^2\theta_{12})$, and thus the kinetic part of the
Hamiltonian, ${\cal H}_\mathrm{kin}$. The role
of solar neutrino data is then to check that the overall amplitude
$a_\mathrm{MSW}$ of the interaction energy difference $V$ in the
dynamical term ${\cal H}_\mathrm{dyn}$
 is consistent with the standard
electroweak model ($a_\mathrm{MSW}=1$).

We have thus performed global analyses including both current
solar neutrino data and current (or prospective) KamLAND data,
with $(\delta m^2,\sin^2\theta_{12},a_\mathrm{MSW})$
unconstrained.
In particular, the analysis of current KamLAND data is based on
the binned energy spectrum of reactor neutrino events observed
above 2.6 MeV (54 events) \cite{KamL}. Prospective KamLAND
spectral data have instead been generated, with the same energy
threshold and binning, by assuming either the LMA-I best-fit point
($\delta m^2=7.3\times 10^{-5}$ eV$^2$ and
$\sin^2\theta_{12}=0.315$) or the LMA-II best-fit point ($\delta
m^2=15.4\times 10^{-5}$ eV$^2$ and $\sin^2\theta_{12}=0.300$)
\cite{KLou}, and increased statistics ($5\times 54$ and $10\times
54$ events). The CHOOZ reactor data \cite{CHOO} are also included.

%
\begin{figure}[t]
\vspace{0pt}
\hspace{30pt}
\includegraphics[scale=0.60]{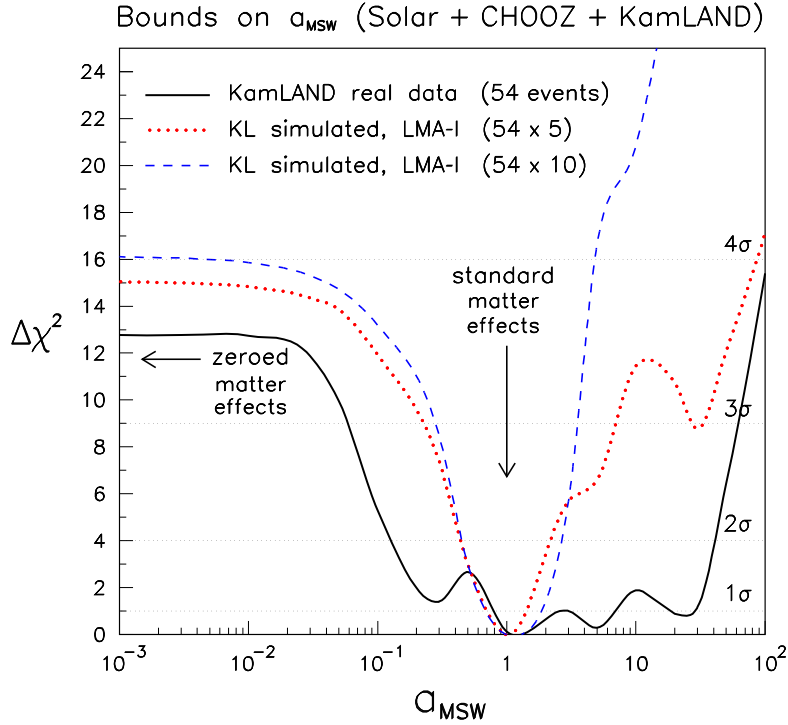}
\vspace{0pt} \caption{\it Bounds on $a_\mathrm{MSW}$ for
unconstrained $(\delta m^2,\,\sin^2\theta_{12})$,  including
current solar and CHOOZ neutrino data, as well as current or
prospective KamLAND data. The solid curve refers to the fit
including {current} KamLAND spectrum data above 2.6 MeV threshold
(54 events), and shows that the hypothetical case of zeroed matter
effects is already disfavored. The other curves refer to
{simulated} KamLAND data, generated by assuming the LMA-I
solution, and statistics increased by a factor of five (dotted
curve) and of ten (dashed curve).}
\end{figure}

%
Figure~7 shows the results of such global fits, in terms of
the function $\Delta \chi^2=\chi^2-\chi^2_{\min}$  for variable
$a_\mathrm{MSW}$ and unconstrained (i.e., minimized away)
mass-mixing parameters. The $n\sigma$ bounds on $a_\mathrm{MSW}$
are then given by $\Delta \chi^2=n^2$. Let us focus first on the
solid curve, which  refers to the fit with {\em current\/} KamLAND
data. It appears that such
curve can already place $>3\sigma$ upper and lower bounds on
$a_\mathrm{MSW}$.  In particular, the hypothetical case of zeroed
matter effects is already disfavored at $\sim 3.5\sigma$, thus
providing an indirect indication in favor of matter effects in the
Sun. The best-fit value of $a_\mathrm{MSW}$ is close to the
standard prediction ($a_\mathrm{MSW}=1$). However,
the overall $\pm 3\sigma$ range
for $a_\mathrm{MSW}$, spanning about three orders of magnitude, is
rather large. The width of this range can be understood by
recalling the following facts: (1) the LMA range of $\delta m^2$
constrained by solar neutrino data, which covers about one decade
\cite{GetM,Last}, can be shifted up or down by shifting
$a_\mathrm{MSW}$ with respect to 1, since the LMA oscillation
physics depends on $V/k\propto a_\mathrm{MSW}/\delta m^2$; (2) the
range of $\delta m^2$ constrained by current terrestrial data
(including KamLAND+CHOOZ), which covers about two decades
\cite{KLou}, is much less affected by $a_\mathrm{MSW}$ variations.
As a consequence, by appropriately shifting $a_\mathrm{MSW}$, it
is possible to overlap the reconstructed ranges of $\delta m^2$
from solar and from reactor data over about $1+2$ decades. When
the overlap sweeps through the degenerate $\delta m^2$ intervals
allowed by KamLAND alone \cite{KLou}, the fit is locally improved,
leading to a ``wavy'' structure in the $\Delta\chi^2$. In
conclusion, although current solar+reactor data strongly disfavor
$a_\mathrm{MSW}=0$ (zeroed matter effects) and provide a best fit
close to $a_\mathrm{MSW}=1$ (standard matter effects), the
presence of other local minima in the $\Delta\chi^2$ function, as
well as the broad $3\sigma$ allowed range for $a_\mathrm{MSW}$, do
not allow to claim a clear evidence of standard matter effects
from current data.

The broken curves in Fig.~7 refer to prospective KamLAND data,
generated by assuming as true solution the LMA-I best-fit point.
The energy threshold, the binning, and the systematic
uncertainties are assumed to be the same as for the current
KamLAND data. The dotted (dashed) curve refers to a number of
reactor neutrino events five (ten) times larger than the current
statistics. It can be seen that the global fit will
progressively constrain $a_\mathrm{MSW}$ within one decade at $\pm
3\sigma$ and, most importantly, will lead to a marked preference
for $a_\mathrm{MSW}\simeq 1$, which is not yet evident in the
present data. In conclusion, if the LMA-I
solution is the true one, there are good prospect to test
unambiguously the occurrence and size of standard matter effects
in the Sun.

\section{Conclusions}

We have analyzed in detail the current SK atmospheric neutrino
data and and the first K2K spectral data, in order to review the
status of standard $\nu_\mu\to\nu_\tau$ oscillations. We
have then provided updated bounds for the standard oscillation
parameters. In particular, the statistical analysis of
the uncertainties reveals that K2K will lead further progress in
this field, especially through higher-statistics tests of the
low-energy spectrum bins.

Going to solar neutrinos,
the KamLAND experiment has clearly selected the LMA region as the
solution to the solar neutrino problem, and has further reduced
the $(\delta m^2,\sin^2\theta_{12})$ parameter space for active
neutrino oscillations. In the $2\nu$ case, we
find that the post-KamLAND LMA solution appears to be basically
split into two sub-regions, LMA-I and LMA-II. The LMA-I solution,
characterized by $\delta m^2\sim 7\times 10^{-5}$ eV$^2$ and
$\sin^2\theta_{12}\sim 0.3$, is preferred by the global fit. The
LMA-II solution represents the second best fit, at about twice the
value of $\delta m^2$.

In the simplest picture, solar neutrino oscillations depend on the
kinematical parameters $(\delta m^2,\sin^2\theta_{12})$ and on
standard dynamical MSW effects in matter. These effects in
current solar neutrino data are starting to emerge through an
increasingly marked preference for $P_{ee}<1/2$, but still remain
not clearly identified.
In order to quantify statistically the occurrence of MSW effects,
we have introduced a free parameter $a_\mathrm{MSW}$ modulating
the amplitude of the $\nu$ interaction energy difference in the
neutrino evolution equation.
By treating $a_\mathrm{MSW}$ as a continuous parameter, we have
then performed a global analysis including current solar, CHOOZ,
and KamLAND data. The results are encouraging, since upper and
lower bounds on $a_\mathrm{MSW}$ appear to emerge at the
$>3\sigma$ level. In particular, the case of ``zeroed'' matter
effects is significantly disfavored. Moreover, the best-fit is
tantalizingly close to the standard expectations for matter
effects ($a_\mathrm{MSW}=1$).

The situation will improve through higher KamLAND statistics. In both  the
LMA-I and LMA-II cases, it appears possible to reduce the current
uncertainty on $a_\mathrm{MSW}$ by about two orders of magnitude.
In conclusion, the selection of a single solution in the LMA oscillation
parameter space appears to be crucial, before any definite
conclusion can be made on the emerging indications of standard
matter effects in the Sun.


\section{Acknowledgments}

G.L.F. thanks the organizers of
the Conference for the kind ospitality. This work is co-financed by the
Italian Ministero dell'Universit\`a e della Ricerca Scientifica e Tecnologica
(MURST) within the ``Astroparticle Physics'' project.

\end{document}